# Negative Group Velocity Pulse Propagation Through a Left-Handed Transmission Line

Rong Jiang [a]    Zhi-Xun Huang [b]    Jing-Yuan Miao [c]    Xin-Meng Liu [c]

a) Zhejiang University of Media and Communication, Hangzhou 310018, P. R. China

b) Communication University of China, Beijing 100024, P. R. China

c) National Institute of Metrology，Beijing 100013，P.R. China

## Abstract

In this paper, the microwave pulse propagation transferred through a left-handed transmission line using Complementary Omega-Like Structures (COLS) loaded was studied. There was a stop band in transmission from 5.6GHz to 6.1GHz, and the anomalous dispersion was causes in this band. Negative group velocity corresponds to the case in which the peak of the pulse exited before the peak of the incident pulse had entered the sample. The negative group velocity reached $(-0.27c \sim -1.85c)$.

Keywords：Negative Group Velocity, Anomalous Dispersion, Left-Handed Transmission Line

In recent years, researchers substantial was interested in the bizarre phenomena that occurred when the electromagnetic wave propagating through dispersive medium. It was well-known that the group velocities of the electromagnetic wave propagating through anomalous dispersion could exceed the speed of light in free space, and even became negative [1]. That is to say, the peak of the pulse would exit the medium before the peak of the incident pulse entered the medium [2]. The peculiar phenomenon had been investigated by the previous researchers. In 1981, S. Chu and S. Wong [3] first observed the phenomenon of negative group velocity in an absorbing medium. Hereafter, many experiments achieved, for example using atomic gases [4,5], photonic crystals [6], ridge waveguide [7], optical fiber [8,9] and the near field of the antenna [10] obtained the negative group velocity.

When electromagnetic wave propagated through the dispersive medium, the group velocity is $v_g = c/[n + f(dn/df)]$. It was known that when anomalous dispersion was occurred, $\frac{dn}{df} < 0$, and when $\frac{dn}{df}$ is small enough, $\frac{dn}{df} < -\frac{n}{f}$, $v_g < 0$ group velocity was negative. In microwave band to obtain $\frac{dn}{df} < -\frac{n}{f}$ we needed to use complex systems and experiment was very different. Based on that, in order to make experiment system easier, we considered when $\frac{dn}{df}$ was negative, n was also negative, and then, the negative group velocity could be achieved easily. Therefore we used left-handed transmission line whose refractive index was negative to obtain the negative group velocity.

In this paper, Complementary Omega-Like Structures (COLS) was used to load microstrip line [11-13] (shown as in Fig.1.) proceeded the experiment. This complementary structure exhibited the same electromagnetic behavior as complementary split ring resonators [14]. Therefore, a negative permittivity /permeability was obtained by the loops/stems of its complementary structure [12]. In Ref.12 and Ref. 14, the basic unit cell for COLS loaded microstrip line and the lumped element equivalent circuit was shown in Fig.2. Respectively, $R_L$ was conductor losses, L was the inductance of the microstrip line, where $L_1 = L_2 = L/2$. $C_k$ was line capacitance which was electrically coupled from backplane to the microstrip line. $C_\Omega$ and $L_\Omega$ were the capacitance and inductance of omega-like element, respectively.

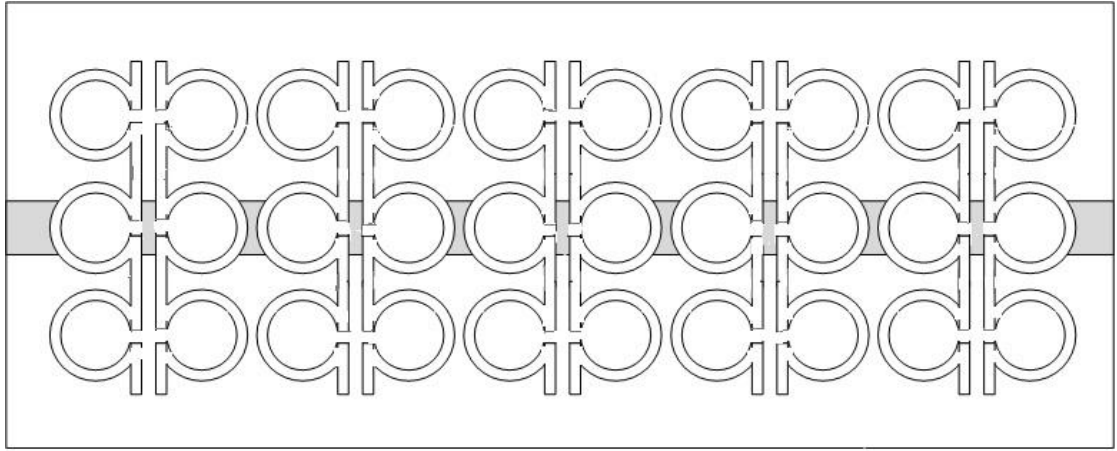

Fig.1. The structure chart of the microstrip transmission line with complementary omega-like structures etched in the ground plane. The gray strip is the microstrip transmission line at the reverse side of printed circuit board.

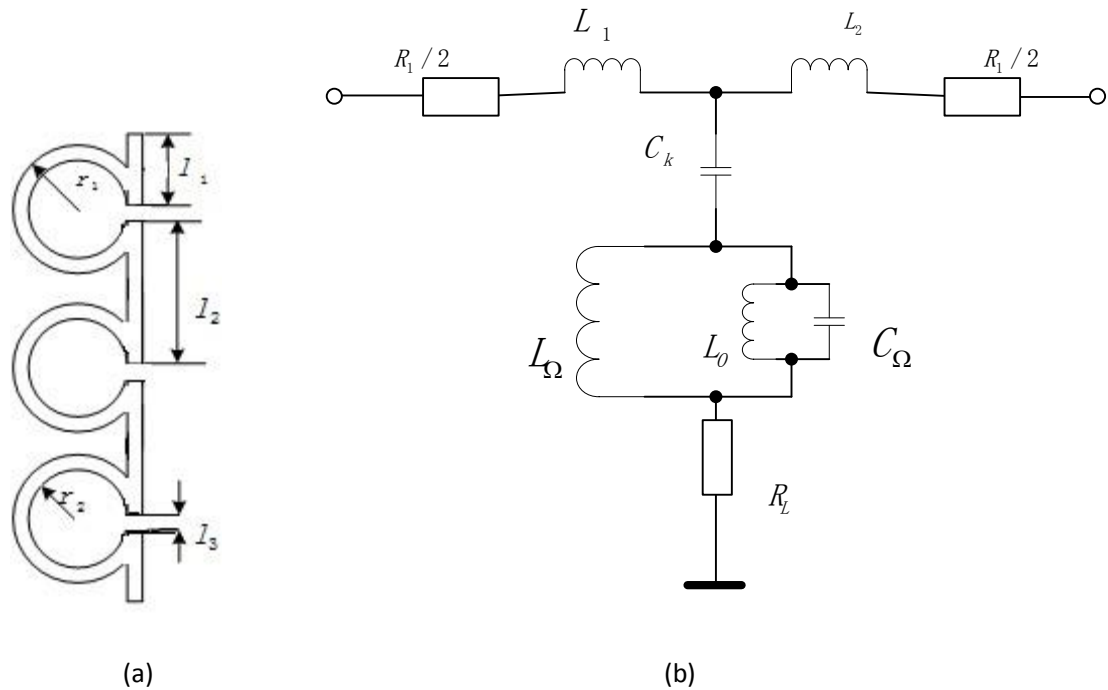

(a) (b)

Fid.2. (a) The structure of basic unit cell for a COLS. (b) equivalent circuit model

By above discussion we find that there are more parameters can be adjusted in the structure of COLS, so the range of anomalous dispersion was found more easily. First, we adjust the size of COLS, and we find that when the size of COLS is relatively small, the anomalous dispersion is strong, group delay is big, but frequency band is narrow. Nevertheless, when the size of COLS is relatively big, the anomalous dispersion is weak, group delay is small, but frequency band is wide. So by calculation the following parameters were chosen shown as in table 1. The structure we chosen was used to find the stop band where anomalous dispersion was occurred. And the phenomenon was observed by experiment used network analyzer(Agilent Technologies E5071C), and experimental set-up was shown in Fig.3. The measurement results was displayed in Fig.4. Then, phase shift was calculated to determine whether exist negative group delay, because the group delay is the negative derivative of the phase shift for frequency $\tau_g = -\frac{d\varphi}{d\omega}$. The result was displayed in Fig.5. In the Fig.5 it is observed that the negative derivative of the phase shift for frequency is negative.

Table 1. (a) This is the parameters of basic unit cell of COLS and the gap between every basic unit. (b) This is the parameters of dielectric slab and the microstrip line

(a)

| basic unit cell of COLS | | | | | |
|---|---|---|---|---|---|
| $r_1$ | $r_2$ | $l_1$ | $l_2$ | $l_3$ | The gap of every unit |
| 0.8mm | 0.5mm | 1.05mm | 2.05mm | 0.2mm | 0.2mm |

(b)

| dielectric slab | | | | microstrip line | |
|---|---|---|---|---|---|
| length | width | height | permittivity | length | width |
| 60mm | 16.8mm | 1.27mm | 8.5 | 60mm | 1.25mm |

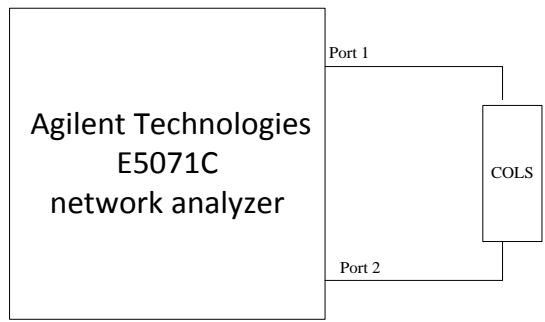

Fig. 3. The experimental set-up of the measurement of anomalous dispersion range

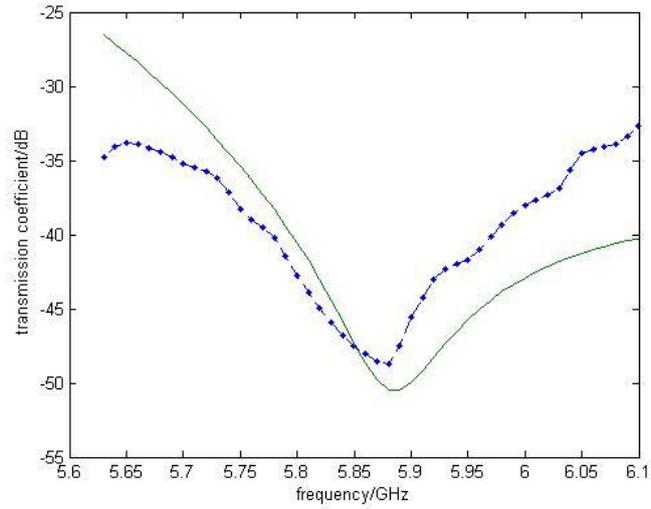

Fig.4. The curve of transmission coefficient. The solid line is calculation results. The dashed line is the experimental results.

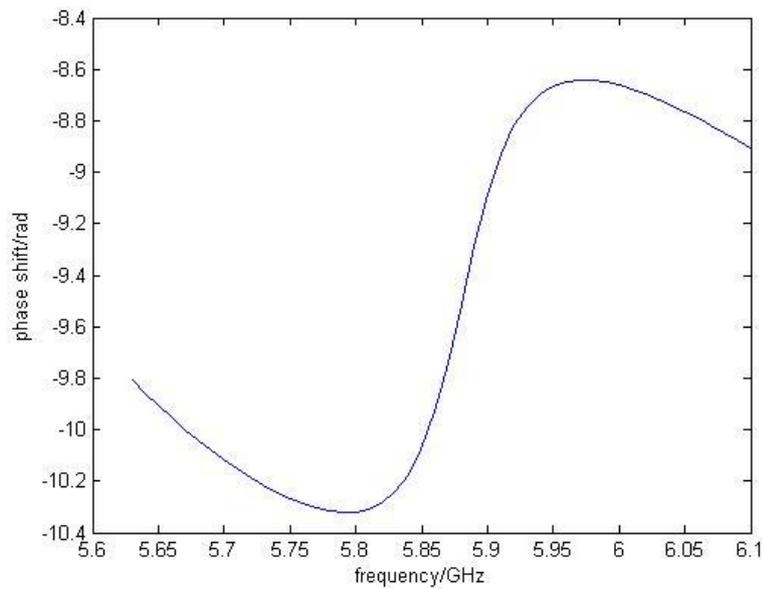

Fig.5. The calculation curve of phase shift

A diagram of experimental system which we used to measure the transit time of microwave pulses along the COLS loaded microstrip line is illustrated in Fig. 6. A sinusoidal carrier wave with a step pulse envelope was launched in the structure using a digital wave generator Agilent E8267D. For carrier frequencies ranging from 5.81 to 6 GHz, the pulse duration was scaled from 5μs. The step pulse was used in order to measure the group delay more easily, because the group delay is very smaller than the pulse width. The amplitude modulated signal fed into a power divider to generate two signals with the same energy. Then two same isolators was connected to two port of power divider in order to prevent reflected signal from interfering the source. Finally, two same test lines were connected to a dual channel digital oscilloscope Tektronix DSA70804B. To observe micro-group-delay we used the real-time demodulation function of oscilloscope. A waveform that the peak of the pulse exits before the peak of the incident pulse has entered the sample at 5.94GHz was presented in Fig.7. Before measuring the group delay, the value time difference of two signal pathway was measured, and that is 0.24 ns. Thereafter, the group delay of the COLS loaded microstrip line was measured. The measurement results are shown in the Table 2. The group delays are all negative from 5.81 to 6 GHz. Through group delay the negative group velocities were obtained. The negative group velocities were $(-0.27c \sim -1.85c)$..

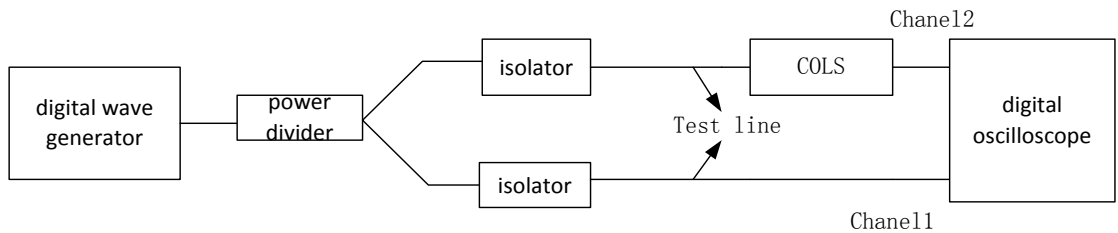

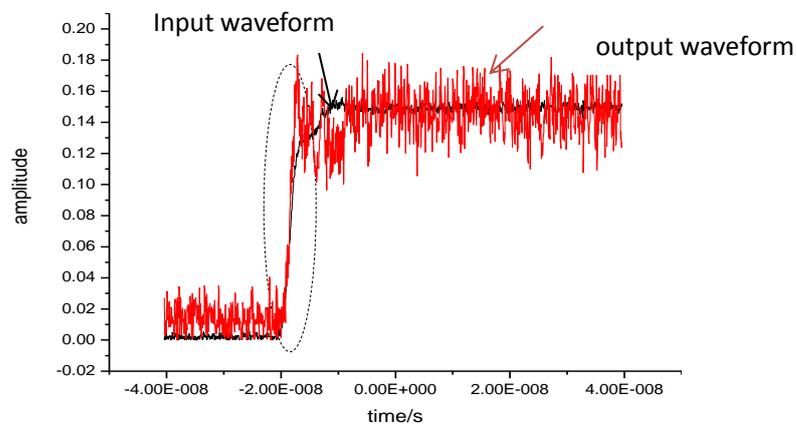

Fig. 6. Schematic diagram of experimental set-up.

(a)

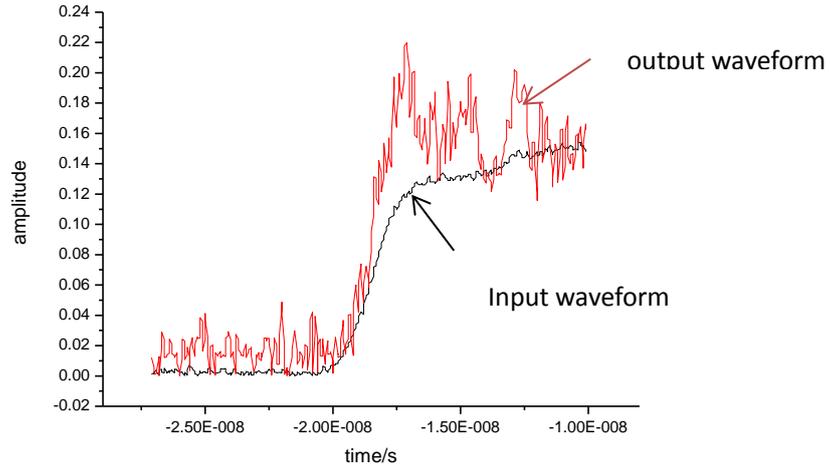

(b)

Fig. 7. (a) is the waveform at 5.94GHz, where the red line is output waveform and black line is input waveform. The output waveform is seen to be advanced in time and to experience distortion and noise in the circle range. (b) is the enlarged picture of circle range. It's obviously shown that output pulse advance

Table 2. Group delay

| frequency GHz | group delay ns | frequency GHz | group delay ns |
|---|---|---|---|
| 5.81 | $-0.259 \sim -0.977$ | 5.91 | $-0.062 \sim -0.698$ |
| 5.82 | $-0.264 \sim -0.267$ | 5.92 | $-0.108 \sim -0.660$ |
| 5.83 | $-0.285 \sim -1.907$ | 5.93 | $-0.261 \sim -1.001$ |
| 5.84 | $-0.520 \sim -1.540$ | 5.94 | $-0.462 \sim -0.859$ |
| 5.85 | $-0.241 \sim -1.12$ | 5.95 | $-0.302 \sim -0.925$ |
| 5.86 | $-0.730 \sim -1.370$ | 5.96 | $-0.071 \sim -0.764$ |
| 5.87 | $-0.127 \sim -0.998$ | 5.97 | $-0.437 \sim -0.743$ |
| 5.88 | $-0.285 \sim -0.840$ | 5.98 | $-0.359 \sim -0.837$ |
| 5.89 | $-0.316 \sim -0.932$ | 5.99 | $-0.263 \sim -0.844$ |
| 5.90 | $-0.239 \sim -0.631$ | 6.00 | $-0.275 \sim -0.700$ |

In conclusion, we have proposed and experimentally demonstrated microwave pulses propagation at negative group velocity by the COLS loaded left-handed transmission line. Compared to the previous negative group velocity experiments[6,7], our experiment scheme is more easily and the accuracy of measurement is higher.

## Acknowledgement

This work is supported by Specialized Research Fund for the Doctoral Program of Higher Education (No. 200800330002).